\documentstyle[sprocl]{article}
\def\bea{\begin{eqnarray}}
\def\eea{\end{eqnarray}}

\input{amssymb.sty}
\def\slaz{\not\!z}
\def\half{\textstyle{\frac{1}{2}}}

\def\C{\mathbb{C}}

\def\cT{\mathcal{T}}\def\Cl{\mathcal{C}\ell}
\def\be{\begin{equation}}\def\ee{\end{equation}}
\def\ra{\longrightarrow}

\begin{document}
\bibliographystyle{unsrt}
\title{LIGHT AS CAUSED NEITHER BY BOUND STATES\\NOR BY
NEUTRINOS\footnote{Submitted in October 1999 to: Valeri V. Dvoeglazov,
editor, {\em Lorentz Group, CPT, and Neutrinos} (World Scientific,
Singapore, 1999 or 2000).}}
\author{KURT JUST}
\address{University of Arizona, Department of Physics\\Tucson, Arizona
85721, USA\\just@physics.arizona.edu}  
\author{KAM-YUEN KWONG}
\address{University of Arizona, Department of Physics\\Tucson, Arizona
85721, USA\\kam@physics.arizona.edu}  

\author{ZBIGNIEW OZIEWICZ\footnote{Supported by el Consejo Nacional de
Ciencia y Tecnolog\'{\i}a (CONACyT) de M\'exico, grant \# 27670 E
(1999-2000). Zbigniew Oziewicz is a member of Sistema Nacional de
Investigadores de M\'exico, Expediente \# 15337.}}
\address{Universidad Nacional Aut\'onoma de M\'exico, Facultad de
Estudios Superiores\\C.P. 54700 Cuautitl\'an Izcalli, Apartado Postal \#
25, Estado de M\'exico\\oziewicz@servidor.unam.mx\\and\\
Uniwersytet Wroc{\l}awski, Poland}
\maketitle\abstracts{Participants of this workshop pursue the old Neutrino 
Theory of Light \cite{perkins} vigorously.  Other physicists have long ago
\cite{wightman} abandoned it, because it lacks \cite{bjorken} gauge
invariance.  In the recent Quantum Induction (QI), all basic Bose fields 
${\mathcal B}^{P}$ are local limits of quantum fields composed 
\cite{qi} of Dirac's
$\Psi$ (for leptons and quarks).  The induced field equations of QI even 
determine all the interactions of those ${\mathcal B}^{P}$.  
Thus a precise gauge
invariance and other physical consequences are unavoidable.  They include the 
absence of divergencies, the exclusion \cite{pauli} of Pauli terms, a 
prediction of the Higgs mass and a `minimal' Quantum Gravity.
  As we find in this paper, however, photons can't be bound states while 
Maxwell's potential $A_{\mu}$ contains all basic Dirac fields except those 
of neutrinos.}

\section*{\bf 1  The Desired Gauge Symmetry}
A topic of this workshop has been the Neutrino Theory of Light; its history 
and present problems have been reviewed by Perkins \cite{perkins}
thoroughly.  Others such as Bjorken \cite{bjorken63} and Wightman 
\cite{wightman} have abandoned it, mainly because they could not attain 
\cite{bjorken} gauge invariance.  This important symmetry poses a serious 
problem as follows.  In the usual theory, Dirac's field ${\Psi}$ and
Maxwell's potential $A_{\mu}$ are basically independent; hence their gauge 
transformations can be postulated separately.  However, when photons are
expected as bound states of basic fermions, should one not ${\it derive}$
the gauge transformation of $A_{\mu}$ from that of ${\Psi}$ ? In the Neutrino
Theory, this question is not even mentioned.

The problem has been solved 
 as a by-product of QI, a version of Quantum Field
Theory where we prevent \cite{qi} the familiar \cite{weinberg} divergencies.  
Since such an unconventional theory cannot be explained briefly, either of 
the following ${\it simplifications}$ must here be adopted: \\
(a) We can treat a mathematical theory similar to QI.  
Then such restrictions may be imposed that we can prove simple results
which formally resemble those of QI. \\
(b) We can explain some results of QI itself completely, but then most proofs
must be omitted for brevity. \\  
Having shown an example of (a) at another \cite{recovery} place, we find (b) 
more suitable for comparing QI with the Neutrino Theory.

In Section 2 we review the postulates of QI and the basic fields entering 
there.  No further postulate is needed, however, for ${\it deriving}$ the
perfect gauge invariance in Section 3.  The composite Bose fields 
${\mathcal B}^{P}$ of QI and their gauge covariant derivatives are 
in Section 4 distinguished from any bound state.
Further extensions mentioned in Section 5 go beyond the expectations
from the Neutrino Theory \cite{perkins} of Light, but fail to satisfy
its old desires.  Thus we are reminded of Quantum Mechanics, which has 
explained much more than Bohr's theory of `planetary' atoms did, but
still cannot yield the `electron orbits' envisioned there.\\

\section*{\bf 2 Fields of Quantum Induction}
With a basically simple, but unorthodox action, we could start from
Feynman's path integral.  For a faster introduction to QI, we (with the
hermitian conjugate ${{\Psi}^{\dagger}}$ and the transpose ${{\Psi}^{T}}$
of the Dirac spinor ${\Psi}$ ) ${\it postulate}$ the anti-commutators\\
\begin{eqnarray}
{[\Psi(x), {\Psi(0)}^{\dagger}]}_{+} \delta(x_0) = \delta(x) 
\  \,   ,\  \,  [\Psi(x),\Psi (0)^T]_{+}\delta(x_0) = 0  
\nonumber  \\
\label{1} \\  
\hbox{and Dirac's equation}\hspace{0.9in}   (i{\not\!\partial}\,
 -\,{\mathcal B})\Psi\, =\, 0 \  . 
\nonumber 
\end{eqnarray}
\newline 
In order to be useful for physics, $\Psi$ must contain the F=24 four-component
spinors $\psi_f$ of all leptons and quarks, although one of these $\psi_f$
suffices for mathematical \cite{recovery} models. 
The Bose field ${\mathcal B}$ 
in (1) is a member of Dirac's Clifford algebra ${\Cl}_{D}$, but also a 
(non-canonical) quantum \cite{jost} field.  Hence it can with a 
suitable basis of ${\Cl}_{D}$ be written

\begin{equation} 
{\mathcal B}\ =\  S^{+}\,+\, i\gamma_5S^- \, +\,{\gamma}^\mu V_\mu^{+}
\, +\, {\gamma}^\mu \gamma_5 V_\mu^-\,+ \,\sigma^{\mu\nu}\cT_{\mu\nu}
\  ,
\label{2}
\end{equation} \\
where $S^\pm$, $V_\mu^\pm$ and $\cT_{\mu\nu}$ act only on the flavors and
colors of $\Psi$. 

In (2) the `tensor potential' $\cT_{\mu\nu}\,=\, -\cT_{\nu\mu}$ must
strictly \cite{pauli} be absent.  Hence we must impose \\ 
\begin{equation}
{\mathcal T}_{\mu\nu}\,=\,0 \qquad  , \hspace{0.5in} \hbox{so that} \qquad  
{\mathcal B}\,=\,S\,+\, \gamma^{\mu} V_{\mu}
\label{3} 
\end{equation}
\newline 
(where $S$, $V_\mu$ are obvious combinations of $S^\pm$, $V_\mu^\pm$ and
$\gamma_5$).
In other words, ${\mathcal B}$ is not spanned by all 16 basis elements of 
Dirac's Clifford algebra ${\Cl}_{D}$, but only by those ten that couple Dirac's
$\Psi$ with $S$ and $V_\mu$ (which contain the observed Higgs and 
Yang-Mills fields).  

This total absence of Pauli terms $\sigma^{\mu\nu} \cT_{\nu\mu}$ from
(2) is necessary whenever postulates similar to (1) are maintained.  Hence
(3) holds as well in the mathematical theory \cite{esposlto} of heat kernels,
but not necessarily in the usual `effective' \cite{weinberg} theory,
where the infinite renormalizations make (1) invalid.  The connections of
tensor potentials ${\mathcal T}_{\mu\nu}\,{\neq}\, 0$ with 
{\it themselves} have been investigated \cite{dvoeglazov} extensively.  
In the literature we cannot find, however, any hint about their 
interactions with Dirac's $\Psi$.
If a tensor potential directly couples only to itself (and perhaps to 
other Bose fields), we evidently cannot reach the conclusion (3);
but how could that hypothetical 
${\mathcal T}_{\mu\nu}\,{\neq}\, 0$ be observed ?

\section*{\bf 3 Implied Gauge Invariance}
Whenever Dirac's $\Psi$ satifies (1) with any Bose field ${\mathcal B}$ 
(even if (3) is violated), this ${\mathcal B}$ can be ${\it recovered}$ 
from a time ordered, bilocal field as the local limit
\begin{eqnarray} {\mathcal B}(x)=8\pi^2\gamma^\mu\lim_{z\rightarrow 0}
T\Psi(x-z){\stackrel{\leftrightarrow}{\partial^x_\mu}}{\overline{\Psi}}(x+z)
{\slaz}^3\  .
\label{4}\end{eqnarray}
Under the restriction to any non-quantized Bose field, its representation
(4) can be derived \cite{recovery} easily.  For clarity, however, we totally
exclude those `purely classical' fields ${\mathcal B}$ 
from QI.  Hence the proof of (4)
requires precautions not explainable briefly; but its technical aspects
remain the same as for non-quantized ${\mathcal B}$.  
Another local limit, which follows
more directly than (4), but also breaks down \cite{brandt} under
infinite renormalizations, is
\begin{eqnarray}
\lim_{z\rightarrow 0}{\not\! z}^{3} b(x,z) \ =\ i \hspace{0.2in}
\hbox{with} \hspace{0.2in}  b(x,z)\,:=\, {(4{\pi})}^{2}T\Psi(x+z) 
\overline\Psi(x-z)
\  .
\label{5}
\end{eqnarray}
Here and in (4), $z=0$ can be approached on any path which does not touch the 
cone $z^{2} = 0$.  

For any member of Dirac's Clifford algebra ${\Cl}_D$, we must now define 
\cite{scadron} the adjoint 
\begin{eqnarray}
{\overline \Gamma} \, :=\, \gamma_0 \Gamma^{\dagger} \gamma_0 \  
\hspace{0.2in} \hbox{for every} \hspace{0.2in} \Gamma\  \in \  {\Cl}_{D}
\   .
\label{6} 
\end{eqnarray}
From (4) and (5), one easily concludes then that the gauge transformation
\begin{eqnarray}
\Psi\ra e^{-i\omega}\Psi\ \hspace{0.4in}
(\  \hbox{hence} \ \quad{\overline{\Psi}}\ra{\overline{\Psi}}
e^{i\overline{\omega}}\ )
\label{7}  
\end{eqnarray}
\begin{eqnarray}
\hbox{implies} \hspace{0.8in}   
{\mathcal B}\ra e^{-i\overline{\omega}}({\mathcal B}-i\not\!\partial)
e^{i\omega}\  .
\label{8}
\end{eqnarray}
Here $\omega$ is a non-quantized matrix, which contains $\gamma_5$ but
no other Dirac matrix (hence $\omega$ acts on flavors, colors and
chiralities, but not on spins or helicities).  We admit 
$\gamma_5\, :=\,  i\gamma^0\gamma^1\gamma^2\gamma^3$ in order to include
chiral transformations; but we make $\omega$ hermitian in order
to exclude conformal mappings (hence the ${\overline {\omega}}$ obeying
$\gamma^{\mu} {\overline {\omega}}\,=\, {\omega} {\gamma}^{\mu}$ simply
equals $\omega$ with $\gamma_5 \longrightarrow -\gamma_5\,$).  From
(8) it is also evident that the gauge transformations (7) are in general 
non-abelian and depend on time and space.  For some readers, (8) will be
more familiar when by (3) it is split into 
\begin{eqnarray}
S\ra e^{-i\overline{\omega}}Se^{i\omega}\  \qquad  \hbox{and}\ \qquad V_\mu\ra 
e^{-i\omega}(V_\mu-i\partial_\mu)e^{i\omega}\qquad .
\label{9}
\end{eqnarray}
or even ${\it linearized}$ in an `infinitesimal' $\omega$.  

We have presented (8) for two reasons: \\
(a) Its derivation from (7) by means of (4) and (5) is ${\it easier}$
than any other proof. Hence it can be left to the reader (just insert (7)
in (4) and apply the product rule of differentiations; the result (8)
follows when (4) and (5) are used).\\
(b) That proof of (8) not only demonstrates 
the implied gauge invariance of (1), but also why
this symmetry has not been achieved with composite photons.  In fact, (4)
does not involve any forces or form factors and not even adjustable 
constants.

\section*{\bf 4 Composite Fields versus States}
The strongest contrast between (4) and any physical `composite' lies in
the following distinction.  For particles to form bound states, they must first
exist themselves.  Hence their own states must be created by operators such as
\begin{eqnarray}
\Psi_f \ :=\ \mbox{\large $\int$} \! f(x) dx \Psi(x) 
\hspace{0.2in} \hbox{with} \hspace{0.2in} f\, \in\,{\mathcal D}\ ,
\label{10}
\end{eqnarray}
obtained by smearing an operator valued distribution \cite{wightm}
with an $f$ from a space ${\mathcal D}$ 
of smooth test functions.  The sharp limit (4), 
however, produces first another local field ${\mathcal B}(x)$, only 
${\it afterwards}$
to be smeared into operators $\int \!f(x) dx {\mathcal B}
(x)$ which create bosons. 

Further consequences of (1) are Dirac induced equations for the Higgs and
Yang-Mills fields contained in (9).  They show that QI prevents the familiar
divergencies very ${\it differently}$ for Dirac's $\Psi$ and the basic
Bose fields in ${\mathcal B}$, namely as follows.  Due to (5), 
the `coincident'
product $\Psi(x) {\overline {\Psi}}(x)$ does not exist; but it is never
needed.  The components of ${\mathcal B}$, however, must form products 
${\mathcal B}^P(x) {\mathcal B}^Q(y)$
which exist (as operator valued distributions) even for $x \equiv  y$.  
Therefore, the quantum fields in ${\mathcal B}$ must be non-canonical, 
just in order to keep $\Psi$ forever canonical.  
Even without proofs, these complete results cannot be explained briefly. 

Splitting (4) as in (3), we also
get $S$ and $V_\mu$ as such local limits. They can be inserted into
\begin{eqnarray}
S_\mu \ :=\  S_{,\mu}+i\left(\overline{V}_\mu S\,-\, SV_\mu\right)
\hspace{0.1in} \hbox{and} \hspace{0,1in}
V_{\mu\nu}\ :=\  V_{\mu,\nu}-V_{\nu,\mu}+i\left[V_\mu,V_\nu\right]
\, .
\label{main11}
\end{eqnarray}
These are called gauge covariant derivatives because they transform
under (9) homogeneously as $S$ (in contrast to $V_\mu$) does.  Then highly 
complicated expressions by $\Psi$ are obtained, involving second derivatives
of $\Psi$, $\overline\Psi$ and terms with $\Psi \overline\Psi\Psi 
\overline\Psi$.  

Very surprisingly, however, there are alternative 
representations of (11) in which not even $\Psi_{, \mu}$ occurs. Instead, 
every component of (11) can ( with a specific constant $c^F_{\mu}$ ) be 
written
\begin{eqnarray}
F(x)=\lim_{z\rightarrow 0}{\langle \ z^\mu T\overline{\Psi}(x-z)
c^F_\mu\Psi(x+z)\  \rangle}_{z^2}\  .
\label{12}
\end{eqnarray}
Here ${{\langle\ldots}\rangle}_{z^2}$ denotes the Lorentz invariant 
average over the 
direction of z, while the matrix 
$c_\mu^F$ acts on the flavors, colors, chiralities
and helicities of $\Psi$. Thus (12) evidently 
is not more complicated than (4), but even much simpler ${\it logically}$.

\section*{\bf 5 Observable Boson States}
Let $\psi_f$ describe a single lepton or quark of the electric charge
$e_f$.  Then the Maxwell component of (12) is
\begin{eqnarray}
F_{\nu\rho}(x)\ =\ const\cdot
\sum_f e_f\lim_{z\rightarrow 0}
{\langle \ {z}^\mu T\overline{\psi}_f(x-z)
\gamma_{\mu\nu\rho}\psi_f(x+z)\  \rangle}_{z^2}
\label{main13}
\end{eqnarray} 
with the `axial' $\gamma_{\mu \nu \rho} \ :=\ \gamma_{[\mu} \gamma_{\nu}
\gamma_{\rho]} \ \in \  {\Cl}_{D}$.  
Hence the state of a single photon, created
with the test function $t^{\nu\rho}\,=\, -t^{\rho\nu}\, \in\, {\mathcal D}$, 
becomes
\begin{eqnarray}
|t\rangle=\mbox {\large $\int$}\!  t^{\nu\rho}(x)dx
\sum_f e_f
\lim_{z\rightarrow 0}{\langle \ {z}^\mu T\overline{\psi}_f(x-z)
\gamma_{\mu\nu\rho}\psi_f(x+z)\ \rangle}_{z^2} |\rangle\ .
\label{main14}
\end{eqnarray}
Here we have used the fact that Maxwell's potential $A_\mu$ is not needed
\cite{stoeger} for ${\it observable}$ photons, although it remains 
indispensible for electromagnetic forces \cite{bohm} and Coulomb 
\cite{steinmann} clouds.  Every pure state with at least one photon follows
from (14), when the Poincare invariant vacuum $|\rangle$ is replaced by the 
most general state in the Hilbert space.  

In order to generalize (14) to any state with at least one observable boson, 
we need only replace (13) 
by other components (12) of (11).  
Not only all Bose fields and their states are
thus recovered, but the Dirac induced field equations of QI also determine 
all their interactions (which for hypothetically 
composite photons are never mentioned).  
Thus QI has achieved much more than one expected from the Neutrino Theory
(even a minimal extension to gravity), but not what there had been desired
orginally.  For instance, the gauge bosons are clearly not bound states,
because neither forces nor form factors occur in (14) or in its 
extensions to other bosons.  Whenever a 
state contains an additional photon, however, this involves due to (14)
all basic Dirac components $\psi_f$ ${\it except}$ those of neutrinos.


\section*{Acknowledgments} For comments we are thankful to 
A. Borowiec, W. Stoeger, E. Sucipto and J. Thevenot.
This work is partially supported by el Consejo Nacional de
Ciencia y Tecnolog\'{\i}a (CONACyT) de M\'exico, grant \# 27670 E
(1999-2000). Zbigniew Oziewicz is a member of Sistema Nacional de
Investigadores de M\'exico.


\end{document}